\documentclass[sort&compress]{elsarticle}
\usepackage[utf8]{inputenc}
\usepackage{amsmath}
\usepackage{amsfonts}
\usepackage{amssymb}
\usepackage{graphicx}
\usepackage{epstopdf}
\usepackage[subnum]{cases}

\newcommand{\br}{\mathbf{r}}
\newcommand{\bx}{\mathbf{x}}
\newcommand{\bp}{\mathbf{p}}
\newcommand{\bE}{\mathbf{E}}
\newcommand{\bB}{\mathbf{B}}
\newcommand{\bA}{\mathbf{A}}
\newcommand{\bv}{\mathbf{v}}
\newcommand{\bP}{\mathbf{P}}
\newcommand{\FF}{{\cal F}}
\newcommand{\GG}{{\cal G}}
\newcommand{\HH}{{\cal H}}

\newtheorem{hypP}{HP}
\newtheorem{hypF}{HF}

\title{Variational formulation of classical and quantum models for intense laser pulse propagation}
\author[gt,amu]{S.A. Berman\corref{cor1}}
\ead{saberman@gatech.edu}

\author[amu]{C. Chandre}

\author[amu]{J. Dubois}

\author[lsu]{F. Mauger}

\author[gt]{M. Perin}

\author[gt]{T. Uzer}

\address[gt]{School of Physics, Georgia Institute of Technology, Atlanta, Georgia 30332-0430, USA}
\address[amu]{Aix Marseille Univ, CNRS, Centrale Marseille, I2M, Marseille, France}
\address[lsu]{Department of Physics and Astronomy, Louisiana State University, Baton Rouge, Louisiana 70803-4001, USA}
\cortext[cor1]{Corresponding author}

\numberwithin{equation}{section}
\begin{document}
\begin{abstract}
We consider the theoretical description of intense laser pulses propagating through gases.
Starting from a first-principles description of both the electromagnetic field and the electron motion within the gas atoms, we derive a hierarchy of reduced models.
We obtain a parallel set of models, where the atomic electrons are treated classically on the one hand, and quantum-mechanically on the other.
By working consistently in either a Lagrangian formulation or a Hamiltonian formulation, we ensure that our reduced models preserve the variational structure of the parent models.
Taking advantage of the Hamiltonian formulation, we deduce a number of conserved quantities of the reduced models.
\end{abstract}
\maketitle
\section{Introduction}
The self-consistent interaction between charged particles and electromagnetic fields is pervasive in physics.
Some common examples include laser-plasma interactions \cite{Esar09}, free electron lasers \cite{Boni87}, and laboratory and astrophysical plasmas \cite{Boyd03}.
Attacking such problems theoretically or even numerically poses a formidable challenge due to the high dimensionality of these systems: The coupling of Maxwell's equations to the charged particle dynamics leads to an infinite-dimensional dynamical system on large spatial scales.
Even if a complete representation of these dynamics were obtainable, it would contain far too much information to allow a clear explanation of the results.
In fact, frequently the results may be explained in terms of simple physical mechanisms which are not substantially affected by the fine details contained in the complete description of the field-particle interaction.
See, for example, the single-wave model for the free electron laser \cite{Boni87,Bach08} and the beam-plasma instability \cite{Tenn94}.
Therefore, we are constantly motivated to seek reduced descriptions which are both numerically tractable and simple enough to permit theoretical analysis of the results and novel experimental predictions.

A powerful and widely-used framework for the reduction of parent models of self-consistent field-particle interaction is the variational formulation \cite{Morr05}.
It consists of casting the first-principles equations either as an action principle or a Hamiltonian system.
For instance, an action principle for the Vlasov-Maxwell equations is given in Ref.~\cite{Low58} and the corresponding Hamiltonian structure is found in Refs.~\cite{Morr80,Mars82}.
Then, simplifying hypotheses for a given problem are incorporated directly into the variational formulation, whether by applying the hypotheses to the action, the Hamiltonian and Poisson bracket \cite{Bach08}, or some combination of the two, as in gyrokinetic theory \cite{Briz07}.
Employing a variational formulation poses several advantages over a reduction performed directly on the equations of motion.
Consistently working in a variational formulation allows the reduced models to preserve conserved quantities possessed by the parent model, avoiding the introduction of unphysical dissipation to the system \cite{Morr05,Cali16}.
Further, variational formulations can provide convenient frameworks for performing arbitrary coordinate transformations \cite{Litt79}.
Lastly, they provide a foundation for the development of specialized numerical schemes which inherently respect the variational structure of the system \cite{Evst13,Stam14,Chen17,Krau13,Krau17,Morr17} and thus may be suitable for accurate long-time integration \cite{McLa92,Tao16}.

In this paper, we consider a variational formulation suitable for describing the propagation of intense, low-frequency laser pulses in gases.
This is the setting for high-harmonic generation (HHG) \cite{Brab00,Gaar08}, terahertz (THz) generation \cite{Amic08,Mart15}, and filamentation \cite{Berg07,Schu17}, to name a few examples.
The parent model which most accurately describes this system is the Maxwell-Schr\"odinger model \cite{Lori07}, which describes the self-consistent interaction between the three-dimensional macroscopic electromagnetic fields and the microscopic wavefunctions describing the atomic or molecular response to the fields.
A first-principles description of the atomic or molecular response is essential for accurately capturing the spatiotemporal evolution of the laser field over experimentally relevant propagation distances \cite{Chri00,Kole13}, which can be hundreds to thousands of times the initial spatial extent of the pulse.
In particular, a quantum or semi-classical description is required to obtain the high-harmonic part of the radiation spectrum with quantitative accuracy \cite{Sand99}.
We have recently shown that a classical description of the atoms self-consistently coupled to the fields can successfully capture the low-frequency part of the spectrum during propagation, as compared with a reduced Maxwell-Schr\"odinger model \cite{Berm18}.
The classical description is also germane for THz generation, where the characteristics of the THz emission may be explained by studying electron trajectories \cite{Mart15}.

Our objective in this work is to use the variational formulations of both the quantum and classical parent models describing intense laser pulse propagation to derive the models employed in Ref.~\cite{Berm18}.
We note that variational formulations of Maxwell-Schr\"odinger models have already been considered for the case of microscopic electromagnetic fields \cite{Chen17,Chen17_2,Masi05,Band13}, though they have not yet been considered for macroscopic fields to the best of our knowledge.
Also, Hamiltonian formulations of reduced laser pulse propagation equations have been found {\it a posteriori}, i.e.\ after reduction from a parent model at the level of the equations of motion \cite{Amir10,Amir16}.

The article is organized as follows.
In Section \ref{sec:parent}, we state the parent models for the classical and quantum systems and the main assumptions that we will incorporate sequentially in order to build a hierarchy of reduced models.
In Section \ref{sec:classical}, we provide the Lagrangian and Hamiltonian derivations of the model with classical dynamics for the particles.
In Section \ref{sec:quantum}, we provide the Lagrangian and Hamiltonian derivations of the model with quantum dynamics for the particles.
Finally, in Section \ref{sec:concl}, we summarize and make some concluding remarks.
Atomic units are used throughout, unless stated otherwise.

\section{Parent model}\label{sec:parent}
Our parent model consists of a classical electromagnetic field interacting with a gas under some reasonable physical assumptions.
For simplicity, we restrict ourselves to the case of single-species single-active-electron (SAE) atomic gases.
The SAE approximation means that we assume the atom consists of a singly-charged ionic core and an electron.
Further, we assume the ions are heavy enough that they may be considered static, at least on the short time scale of the laser pulse.
We also assume the electron motion can be treated in the dipole approximation.
This means the electrons are non-relativistic and move on spatial scales small compared to those of the spatial variations of the electromagnetic field, implying magnetic effects are neglected.
Lastly, we assume a low-density gas such that collisions between electrons and neighboring atoms may be neglected, so each electron only interacts with its parent ion and the macroscopic electric field.

\begin{figure}
\centering
\includegraphics[width=0.9\textwidth]{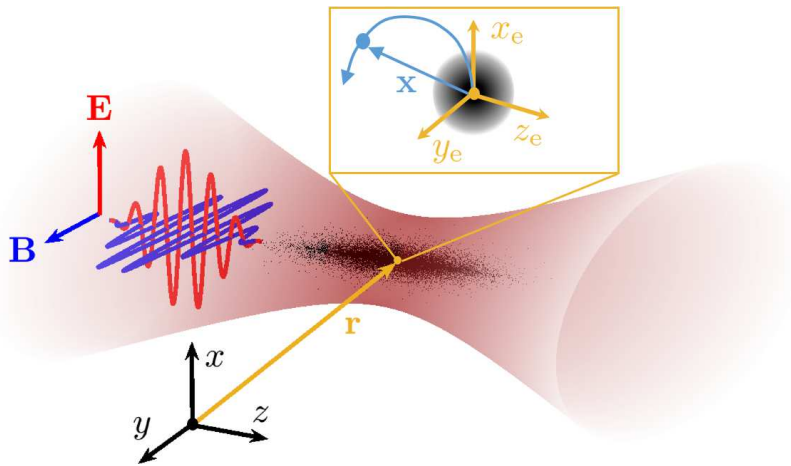}
\caption{Schematic illustrating the typical geometry of an intense, linearly-polarized laser pulse propagating through an atomic gas. The incident pulse is on the left and it propagates in the positive $z$ direction. The intensity profile of the focused laser beam is shown in dark red. The gas atoms are the black dots distributed around the focus of the laser beam. $\br$ is the macroscopic coordinate, such that the laser electric field is $\bE(\br,t)$ (in red), the magnetic field is $\bB(\br,t)$ (in blue), and the gas density is $\rho(\br)$. The inset shows that at each point $\br$, a microscopic coordinate $\bx$ is attached, giving the position of the electron of an atom located at $\br$. }\label{fig:schematic}
\end{figure}

We define the dynamical variables and coordinate systems of our parent model, illustrated in Fig.~\ref{fig:schematic}, as follows.
The dipole approximation leads naturally to a separation of length scales into a macroscopic scale and a microscopic scale.
The coordinate of the macroscopic scale, $\br = (x,y,z)$, gives the position of an arbitrary point in the gas. 
Meanwhile, the coordinate of the microscopic scale, $\bx = (x_{\rm e},y_{\rm e},z_{\rm e})$, gives the position of the electron relative to the ionic core of an atom.
The dynamical variables for the electromagnetic field are $\bE(\br,t)$ and $\bB(\br,t)$.
The atomic number density is $\rho(\br)$, which is time-independent due to the assumption of static ions.
For the quantum model, the dynamical field variable for the particles is the electronic wavefunction $\psi(\bx,t;\br)$ of an atom located at $\br$.
For the classical model, the dynamical field variable for the particles is $\bx(\br,t;\bx_0,\bv_0)$, which gives the position $\bx$ of the electron of an atom located at $\br$ at time $t$.
The labels $\bx_0$ and $\bv_0$ define the initial position and velocity, respectively, i.e.\ $\bx(\br,0;\bx_0, \bv_0)=\bx_0$ and  $\dot{\bx}(\br,0;\bx_0, \bv_0)=\bv_0$.

The equations of motion for the electromagnetic fields are Maxwell's equations, which read \cite{Lori07}
\begin{subequations}\label{eq:Maxwell}
\begin{align}\label{eq:MaxAmpere}
& \dot{\bE} =  c^2 \nabla \times \bB - 4\pi \dot{\bP}, \\ \label{eq:MaxLenz} 
& \dot{\bB} = - \nabla \times \bE, \\ \label{eq:MaxGauss}
& \nabla \cdot \bE = - 4\pi \nabla \cdot \bP, \\ \label{eq:MaxDivB}
& \nabla \cdot \bB = 0.
\end{align}
\end{subequations}
Here, $\bP$ is the macroscopic polarization which, in this case, can be expressed as
\begin{equation}\label{eq:Polarization}
\bP(\br,t) = -\rho(\br) \overline{\bx}(\br,t),
\end{equation}
where $\overline{\bx}$ is the ensemble-averaged electron position of the atoms located near $\br$ at time $t$.
That is, we average over the large number of atoms $\rho(\br)\mathrm{d}^3 \br$ contained in a small volume $\mathrm{d}^3 \br$ around $\br$ (see Fig.~\ref{fig:schematic}).
In order to solve Eqs.~\eqref{eq:Maxwell}, we must specify a microscopic model for the response of a single atom to the field, which allows one to determine $\overline{\bx}$ from $\bE$.

Typically, a quantum model is employed to obtain the single-atom response to the field, as in the Maxwell-Schr\"odinger model \cite{Lori07}.
The equations are
\begin{subequations}\label{eq:QuantParent}
\begin{align}\label{eq:ParentSchrod}
& i \dot{\psi} = - \frac{1}{2}\nabla^2_\bx \psi + \big[V(\bx) + \bE(\br,t) \cdot \bx \big] \psi, \\ \label{eq:ParentExpecVal}
& \overline{\bx}(\br,t) = \int \bx \, |\psi(\bx,t; \br)|^2 \,\,\mathrm{d}^3\bx.
\end{align}
\end{subequations}
Here, Eq.~\eqref{eq:ParentSchrod} is the Schr\"odinger equation for the wave function $\psi(\bx,t;\br)$.
The ion-electron interaction is described by an effective potential $V$, such as the soft-Coulomb potential $V(\bx)=-(|\bx|^2+1)^{-1/2}$ \cite{Java88,Beck12,Berm18}.
The use of the dipole approximation is evident from the fact that $\bE$ depends on the macroscopic coordinate $\br$, but not on the microscopic coordinate $\bx$.
The latter is the sole degree of freedom of the electron, as illustrated in Fig.~\ref{fig:schematic}.
Equation \eqref{eq:ParentExpecVal} is the quantum expectation value of the observable $\bx$, with the integration carried out over all $\bx \in \mathbb{R}^3$.
We assume that the ensemble of atoms located at $\br$ is initially in a pure state $\psi(\bx,0;\br)=\psi_0(\bx)$, i.e. each of the atoms is in the same initial state, so we do not need to consider a density matrix to describe the ensemble.
Note that, in principle, $\psi_0$ could also depend on $\br$, but we choose to make it independent of $\br$ so that the initial state of the atoms is uniform.
Thus, the expectation value $\overline{\bx}$ in Eq.~\eqref{eq:ParentExpecVal} is indeed the ensemble-averaged electron position of the atoms at $\br$.
Together, Eqs.~\eqref{eq:Maxwell}-\eqref{eq:QuantParent} constitute the Maxwell-Schr\"odinger model, which we refer to here as the parent quantum model.

The quantum model is particularly effective in the context of HHG, where it accurately describes the evolution of the high harmonic radiation during propagation \cite{Lori07,Lori12}.
However, the description of the electron dynamics in terms of a time-dependent wave function lacks the intuitive and very relevant dynamical picture provided by the underlying classical electron trajectories \cite{Cork93,Kula93,Kamo14}.
To address this issue, one option is to use purely classical models for the electron dynamics \cite{Sand99, Berm18, Band90,Band92,Both09,Uzdi10}, in which the Schr\"odinger equation is replaced with the corresponding classical equations of motion for $\bx(\br,t;\bx_0, \bv_0)$, and the quantum expectation value is replaced with a classical average over an ensemble of initial conditions.
The ensemble of initial conditions is typically chosen to capture as much as possible the effects one would observe in a quantum description, such as wavepacket spreading.
The corresponding equations are
\begin{subequations}\label{eq:ClassicalParent}
\begin{align}\label{eq:ParentNewton}
& \ddot{\bx} = -\nabla_\bx V(\bx) - \bE(\br,t), \\ \label{eq:ParentEnsAve}
& \overline{\bx}(\br,t) = \int \bx(\br,t;\bx_0, \bv_0) f_0(\bx_0,\bv_0) \mathrm{d}^3 \bx_0 \mathrm{d}^3 \bv_0.
\end{align} 
\end{subequations}
The dipole approximation is reflected in the same way in Eq.~\eqref{eq:ParentNewton} as in the quantum case.
The integral in Eq.~\eqref{eq:ParentEnsAve} is carried out over all $(\bx_0,\bv_0) \in \mathbb{R}^3\times \mathbb{R}^3$.
Meanwhile, $f_0$ is the probability distribution function to find an electron with the given initial conditions.
By averaging with respect to $f_0$ in Eq.~\eqref{eq:ParentEnsAve}, we obtain the ensemble-averaged position of the electron relative to the ion $\overline{\bx}$, which allows us to obtain the polarization using Eq.~\eqref{eq:Polarization}.
As in the quantum case, we assume the initial state of the atoms $f_0$ is independent of $\br$.
Together, Eqs.~\eqref{eq:Maxwell}, \eqref{eq:Polarization}, and \eqref{eq:ClassicalParent} constitute the parent classical model.

\begin{table}
\begin{tabular}{|p{5.5cm} | p{5.5cm} |}
\hline
\begin{hypP}\label{h:SAE}
SAE approximation
\end{hypP}
&
\begin{hypF}\label{h:plane}
Electromagnetic fields solely depend on propagation coordinate $z$
\end{hypF}
\\
\begin{hypP}\label{h:ions}
static ions
\end{hypP}
&
\begin{hypF}\label{h:electrostatic}
$z$-component of the electric field is negligible
\end{hypF}
\\
\begin{hypP}\label{h:dipole}
dipole approximation
\end{hypP}
&
\begin{hypF}\label{h:LP}
linearly polarized fields
\end{hypF}
\\
\begin{hypP}\label{h:planeP}
Particle fields solely depend on propagation coordinate $z$
\end{hypP}
&
\begin{hypF}\label{h:back}
backward-propagating waves are negligible
\end{hypF} 
\\
\begin{hypP}\label{h:ePS}
reduced electron phase space
\end{hypP}
&
\\
\hline
\end{tabular}
\caption{The various hypotheses underlying our reduced models. The left column contains hypotheses on the response of the particles to the field (HP), while the right column contains hypotheses on the fields themselves (HF).}\label{tab:Hyp}
\end{table}

For both parent models, it is not possible to obtain the microscopic dynamics nor the macroscopic dynamics analytically.
Hence, one must resort to numerical simulations.
However, the computational cost of simulating the parent models is immense due to the multiscale nature of the problem.
For instance, numerically solving the quantum model requires one to obtain the solution $\psi(\bx,t;\br)$ of Eq.~\eqref{eq:ParentSchrod} at every time step  for all $\br$ and $\bx$ in the computational domain.
In the present day, such a feat can only be accomplished using super-computers \cite{Lori07,Lori12}.
Thus, reduced models with smaller computational requirements are highly desirable.

In the following sections, we will build a hierarchy of reduced models stemming from the two parent models by sequentially incorporating the hypotheses outlined in Table \ref{tab:Hyp}.
The parent models already incorporate hypotheses HP \ref{h:SAE}-\ref{h:dipole}.
In this article, we choose to focus on the simplification of the field part of the equations, because any dimensional reduction on the macroscopic scale automatically results in fewer computationally-costly microscopic computations \cite{Kole04}.
Furthermore, reductions on the particle dynamics tend to be less general and rely on specific hypotheses particular to certain sets of field parameters.
Such reductions typically do not interfere with the structure of the self-consistent interaction between the field and the particles, so they may be built on top of the models we arrive at in this article.
In fact, because we focus on reducing the electromagnetic fields, the models we obtain may be readily generalized to other types of gas species or condensed phase systems, simply by specifying the appropriate microscopic model for the reponse of the medium to the fields \cite{Kole04}.

\section{Classical reduced models}\label{sec:classical}
\subsection{Lagrangian formulation}\label{sec:LagClassical}
The classical model, Eqs.~\eqref{eq:Maxwell}, \eqref{eq:Polarization}, and \eqref{eq:ClassicalParent}, admits a Lagrangian formulation.
First, we introduce the electromagnetic potentials, the scalar potential $\varphi(\br,t)$ and the vector potential $\bA(\br,t)$, from which the electric and magnetic fields are obtained as 
\begin{subequations}\label{eq:potenials}
\begin{align}\label{eq:potE}
\bE & = -\nabla \varphi - \dot{\bA}, \\ \label{eq:potB}
\bB & = \nabla \times \bA.
\end{align}
\end{subequations}
Now, we define the action functional $\mathcal{A}[\bx(\br,t),\varphi(\br,t),\bA(\br,t)]$ as
\begin{subequations}\label{eq:Lagrangian}
\begin{align}
& \mathcal{A}[\bx,\varphi,\bA] = \int (\mathcal{L}_\mathrm{P} + \mathcal{L}_\mathrm{EM} ) \mathrm{d}t, \\ \label{eq:LagParticle}
& \mathcal{L}_\mathrm{P} = 4\pi \int \rho \left[ \frac{|\dot{\bx}|^2}{2} - V(\bx) + \bx \cdot \nabla \varphi - \dot{\bx} \cdot \bA \right] \mathrm{d}\mu \mathrm{d}^3\br, \\
& \mathcal{L}_\mathrm{EM} = \frac{1}{2} \int \left(|\nabla \varphi + \dot \bA|^2 - c^2 |\nabla \times \bA|^2 \right) \mathrm{d}^3\br,
\end{align}
\end{subequations}
where we have introduced the notation $\mathrm{d}\mu = f_0(\bx_0,\bv_0) \mathrm{d}^3\bx_0 \mathrm{d}^3 \bv_0$.
Hence, the Lagrangian is decomposed into a particle Lagrangian $\mathcal{L}_\mathrm{P}$ and an electromagnetic Lagrangian $\mathcal{L}_\mathrm{EM}$, in a manner similar to the Low Lagrangian for the Vlasov-Maxwell equations \cite{Low58}.

Imposing $\delta \mathcal{A} = 0$, i.e.\ $\mathcal{A}_\mathcal{\bx} = \mathcal{A}_\mathcal{\varphi} =  \mathcal{A}_{\bA} = 0$, and applying Eqs.~\eqref{eq:potenials} yields the parent model equations \eqref{eq:Maxwell} and \eqref{eq:ClassicalParent}.
The subscript denotes the functional derivative, defined for a functional $\FF[f({\bf z})]$ of a function $f$ on an $n$-dimensional domain by
\begin{equation*}
\FF[f({\rm z})+ \varepsilon \delta f({\rm z})] - \FF[f({\rm z})] = \varepsilon \int \FF_f \delta f {\rm d}^n{\bf z} + {\cal O}(\varepsilon^2).
\end{equation*}
The first of these equations yields Eq.~\eqref{eq:ParentNewton}, when one requires that $\mathcal{A}_{\bx} = 0$ for an arbitrary $f_0$.
The second yields
\begin{equation}\label{eq:LagGauss}
-(\nabla^2 \varphi + \nabla\cdot \dot{\bA}) = 4\pi \nabla \cdot \left(\rho \int \bx \,\mathrm{d}\mu\right).
\end{equation}
This is equivalent to Gauss' Law, Eq.~\eqref{eq:MaxGauss}, upon applying Eqs.~\eqref{eq:potE}, \eqref{eq:ParentEnsAve}, and \eqref{eq:Polarization}.
Similarly, the third yields
\begin{equation*}
-(\nabla \dot{\varphi} + \ddot{\bA}) = c^2 \nabla \times (\nabla \times \bA) + 4 \pi \rho \int \dot{\bx} \,\mathrm{d}\mu,
\end{equation*}
which is equivalent to the Maxwell-Amp\`{e}re equation \eqref{eq:MaxAmpere}.
Meanwhile, Eqs.~\eqref{eq:MaxLenz} and \eqref{eq:MaxDivB} are automatically satisfied by virtue of Eqs.~\eqref{eq:potenials}.

We remark that the existence of a variational priniciple is a fundamental property of the system.
That is, it is not a consequence of choosing a particular set of dynamical variables, such as the electromagnetic potentials.
For example, one may write an action principle using the electric and magnetic fields themselves as dynamical variables, instead of the potentials, but the formulation is more complicated as it requires constrained variations.
Thus, we use the electromagnetic potentials for convenience.

Now, we begin making assumptions and approximations appropriate to typical experimental situations in order to obtain a hierarchy of reduced models.
In experiments, the radiation is generated by the propagation of spatially localized laser pulses with a given polarization through the gas.
The spatial localization comes from the focusing of the laser beam, which typically leads to a Gaussian intensity profile with cylindrical symmetry about the propagation axis (see Fig.~\ref{fig:schematic}).
As such, the vacuum field intensity only depends on the propagation coordinate $z$ and the distance from the propagation axis $|\br_\perp|$, where $\br_\perp = (x,y)$.
The time-dependent electric field then consists of the product of a spatial envelope due to focusing, a  temporal envelope due to finite pulse duration, and an oscillation at the carrier frequency of the laser $\omega_L$.
Notably, the focusing leads to a $z$-dependent maximum amplitude of the field as well as a $z$-dependent carrier-envelope phase, known as the Gouy phase shift \cite{Zang13}, even in vacuum.
Focusing effects can be described efficiently within the paraxial approximation, which is valid for laser beams that are not too tightly focused.
However, even within the paraxial approximation, the dimensionality of the electromagnetic fields is not substantially reduced.
The fields still depend on two spatial coordinates, $r_\perp$ and $z$.
Furthermore, even though it is true that the fields are dominated by the transverse components, i.e.\ the $\br_\perp$ direction, Gauss' Law requires that they have a longitudinal component as well \cite{Zang13}.
Hence, the fields are still three-dimensional vectors in the case of arbitrary laser polarization, and still two-dimensional in the simpler (and very common) case of linear polarization.
Lastly, the paraxial approximation inherently assumes backward-propagating waves are negligible \cite{Kole13}, but this is not actually the case if the gas density is high enough \cite{Shon00}.

\subsubsection{Plane-wave fields (HP \ref{h:SAE}-\ref{h:planeP}, HF \ref{h:plane})}
Here, we make strong assumptions on the fields to bypass these difficulties.
Namely, we assume that the fields' only spatial dependence is on $z$, invoking hypotheses HF \ref{h:plane} and HP \ref{h:planeP} of Table \ref{tab:Hyp}.
We take the fields to be plane waves of the form
\begin{subequations}\label{eq:fieldAssump}
\begin{align}
\rho & = \rho(z), \\ \label{eq:fieldAssump_x}
\bx & = \bx(z,t;\bx_0,\bv_0), \\
\varphi & = \varphi(z,t), \\
\bA & = \bA_\perp(z,t) = A_x(z,t)\hat{\mathbf{x}} + A_y(z,t)\hat{\mathbf{y}}.
\end{align}
\end{subequations}
In writing the vector potential, we have employed the radiation gauge $\nabla \cdot \bA=0$ which, for $z$-dependent fields, becomes $\partial_z A_z = 0$.
While this technically allows $A_z$ to be a function of time, we have chosen $A_z=0$ to avoid the presence of a uniform time-dependent electric field.
As is evident from Eq.~\eqref{eq:LagGauss}, the main appeal of adopting the radiation gauge is that all the space charge effects, i.e.\ those due to a spatially nonuniform charge distribution, are encoded in $\varphi$.

Inserting Eqs.~\eqref{eq:fieldAssump} into Eqs.~\eqref{eq:Lagrangian}, we obtain the Lagrangians of the reduced system as
\begin{subequations}\label{eq:Model_1D_fields}
\begin{align} \label{eq:Mod_1D_fields_Lp}
& \mathcal{L}_\mathrm{P} = 4\pi  \int \rho \left[ \frac{|\dot{\bx}|^2}{2} - V(\bx) + z_{\rm e} \, \partial_z \varphi - \dot{\bx}_\perp \cdot \bA_\perp \right] \mathrm{d}\mu \mathrm{d}z, \\ \label{eq:Mod_1D_fields_Lem}
& \mathcal{L}_\mathrm{EM} = \frac{1}{2} \int \left[(\partial_z \varphi)^2 + |\dot{\bA}_\perp|^2  - c^2  |\partial_z \bA_\perp|^2 \right]\mathrm{d}z.
\end{align}
\end{subequations}
By taking variations with respect to the present variables, we obtain the equations of motion
\begin{subequations}\label{eq:EOM1} 
\begin{align} \label{eq:EOM1Particle}
& \ddot{\bx} = -\nabla_\bx V(\bx) + (\partial_z \varphi) \hat{\mathbf{z}} + \dot{\bA}_\perp, \\\ \label{eq:EOM1Gauss}
& \partial^2_{z} \varphi = -4\pi \partial_z(\rho \overline{z}_{\rm e}), \\ \label{eq:EOM1Wave}
& c^2 \partial^2_{z} \bA_\perp - \ddot{\bA} =4 \pi \rho \dot{\overline{\bx}}_\perp. 
\end{align}
\end{subequations}
Lagrangians \eqref{eq:Model_1D_fields} and the corresponding equations of motion constitute the first reduced model we have obtained, including assumptions HP \ref{h:SAE}-\ref{h:planeP} and HF \ref{h:plane} in Table \ref{tab:Hyp}.
Together, they describe the self-consistent dynamics of arbitrarily polarized transverse electromagnetic plane waves and longitudinal space charge waves propagating through a classical atomic gas.
Treating the fields as plane waves corresponds to assuming that the laser beam is loosely focused and the focusing can be neglected altogether.
Admittedly, this hypothesis is rarely met in experiments, but it is the key to obtaining a substantial dimensional reduction from the parent model.
Later, we shall show how one can reintroduce some of the focusing effects externally.

\subsubsection{Transverse, linearly-polarized fields (HP \ref{h:SAE}-\ref{h:planeP} and HF \ref{h:plane}-\ref{h:LP})}
Next, we make an assumption which allows us to remove the scalar potential $\varphi$.
Specifically, we assume that the $E_z$ is negligible (HF \ref{h:electrostatic}).
Since  $E_z = -\partial_z \varphi$, this is equivalent to assuming $\partial_z \varphi$ is negligible.
This is easily justified for atoms, where the symmetry of the ionic potential $V$ and initially purely transverse electric field guarantee that $\overline{z}_{\rm e}(z,t)=0$ for all $z$ and $t$.
In turn, this makes $\partial_z \varphi (z,t)=0$ by virtue of Eq.~\eqref{eq:EOM1Gauss}.
Thus, the presence of $\partial_z \varphi$ makes no difference for the atomic response to the fields.
However, $\overline{z}_{\rm e}$ need not be zero for anisotropic media such as aligned molecules, where $V$ is  asymmetric.
In that case, it may still be reasonable to neglect $\partial_z \varphi$ because it may be very small.
In particular, this hypothesis can always be met for a small enough density $\rho$, since under hypothesis HF \ref{h:plane}, the only possible source for the longitudinal component of the electric field is the radiation of the particles.

Assuming $\partial_z \varphi$ is negligible, we drop the terms containing it from the Lagrangian.
Since these are the only places where $\varphi$ appears, it is eliminated as a dynamical field by hypothesis HF \ref{h:electrostatic}.
This leaves us with the Lagrangians
\begin{subequations}\label{eq:Model_1D_wave}
\begin{align} \label{eq:Mod_1D_wave_Lp}
& \mathcal{L}_\mathrm{P} = 4\pi  \int \rho \left[ \frac{|\dot{\bx}|^2}{2} - V(\bx) - \dot{\bx}_\perp \cdot \bA_\perp \right] \mathrm{d}\mu \mathrm{d}z, \\ \label{eq:Mod_1D_wave_Lem}
& \mathcal{L}_\mathrm{EM} = \frac{1}{2} \int \left[|\dot{\bA}_\perp|^2 - c^2 |\partial_z\bA_\perp|^2 \right] \mathrm{d}z.
\end{align}
\end{subequations}
Above we have the next reduced model in the hierarchy, incorporating hypotheses HP  \ref{h:SAE}-\ref{h:planeP} and HF \ref{h:plane}-\ref{h:electrostatic} of Table \ref{tab:Hyp}.
The equations of motion are Eqs.~\eqref{eq:EOM1Particle} and \eqref{eq:EOM1Wave}, with the $\partial_z \varphi$ term in Eq.~\eqref{eq:EOM1Particle} omitted.

In what follows, we will restrict our attention to linearly-polarized waves for simplicity, though our subsequent reductions apply equally well in the case of arbitrary polarization.
We invoke assumption HF \ref{h:LP} and take $A_y=0$.
In this case, the equations of motion are
\begin{subequations}\label{eq:EOM2} 
\begin{align} \label{eq:EOM2_particle}
& \ddot{\bx} = -\nabla_\bx V(\bx) + \dot{A}\hat{{\bf x}}, \\ \label{eq:EOM2_wave}
& c^2 \partial^2_{z} A - \ddot{A} = 4 \pi \rho \dot{\overline{x}}_{\rm e},
\end{align}
\end{subequations}
where we have omitted the $x$-subscript on $A$.
Thus, in Eq.~\eqref{eq:EOM2_particle} we have the equation of motion for an atomic electron driven by the electric field $E=-\dot{A}$ of the electromagnetic wave.
Meanwhile, in Eq.~\eqref{eq:EOM2_wave} we have a 1D wave equation for $A$, with the $x$-component of the current density of the atoms on the right-hand side as a source term.

\subsubsection{Moving frame}
Before proceeding to the next reduced model, we perform a change of coordinates into a moving frame which is better suited to the analysis of propagating laser pulses.
Because the laser pulse moves at nearly the speed of light through the gas, it is convenient to change the coordinates of the fields to $\xi = z$ and $\tau = t - z/c$.
This leads to an equivalent action $\widetilde{\cal A}[\widetilde{\bx}(\xi,\tau),\widetilde{A}(\xi,\tau)] = \int ( \widetilde{{\cal L}}_{\rm P} + \widetilde{{\cal L}}_{\rm EM} ) {\rm d}\tau$, defined such that $\widetilde{{\cal A}}[\widetilde{\bx},\widetilde{A}] = {\cal A}[\bx,A]$.
In particular, the new arguments of $\widetilde{{\cal A}}$ are defined such that
\begin{subequations}
\begin{align*}
\widetilde{\bx}(\xi,\tau;\widetilde{\bx}_0,\widetilde{\bv}_0) & = \bx(\xi,\tau + \xi/c;\bx_0,\bv_0), \\
\widetilde{A}(\xi,\tau) & = A(\xi,\tau + \xi/c).
\end{align*}
\end{subequations}
Here, we have introduced the positions and velocities of the atomic electrons at $\tau=0$, when the laser pulse arrives to their location $\xi$ along the propagation direction,
\begin{subequations}
\begin{align*}
\widetilde{\bx}_{0} & = \bx(\xi,\xi /c; \bx_{0}, \bv_0), \\
\widetilde{\bv}_{0} & = \dot{\bx}(\xi,\xi/c;\bx_{0}, \bv_{0}),
\end{align*}
\end{subequations}
which play the role of initial conditions in the moving frame.
We also define the distribution of electron initial conditions when the pulse arrives as $\widetilde{f}_0(\widetilde{\bx}_{0},\widetilde{\bv}_{0}) = f_0(\bx_{0},\bv_{0})$.
Applying the chain rule, we obtain the new Lagrangians
\begin{subequations}\label{eq:Model_moving}
\begin{align} \label{eq:Mod_mov_Lp}
& \mathcal{L}_\mathrm{P} = 4\pi \int \rho\left[ \frac{|\partial_\tau \bx|^2}{2} - V(\bx) - \partial_\tau x_{\rm e} A \right] \mathrm{d}\mu \mathrm{d}\xi, \\ \label{eq:Mod_mov_Lem}
& \mathcal{L}_\mathrm{EM} = \int \left[ c \partial_\tau{A}\,\partial_\xi A - \frac{c^2}{2}(\partial_\xi A)^2\right] \mathrm{d}\xi,
\end{align}
\end{subequations} 
where we have omitted the tildes over the new Lagrangians and the new field variables.
The equations of motion for $\bx$ have the same form in the moving frame because the microscopic coordinates are unaffected by the moving-frame transformation.
The equation for $A$ becomes
\begin{equation}\label{eq:EOM_Amoving}
c^2 \partial^2_{\xi} A - 2 c \partial_\xi \partial_\tau{A} = 4 \pi \rho \partial_\tau{\overline{x}}_{\rm e}.
\end{equation}
We observe that, in these coordinates, the equation for $A$ has become first-order in time $\tau$, though it is still second-order in space $\xi$.

\subsubsection{Unidirectional approximation (HP \ref{h:SAE}-\ref{h:planeP} and HF \ref{h:plane}-\ref{h:back})}\label{sec:LagUni}
By making a certain hypothesis on the derivatives of $A$, we remove the second-order derivative and obtain the next reduced model in our hierarchy.
Looking at Lagrangian \eqref{eq:Mod_mov_Lem}, we observe that if $|\partial_\xi A| \ll |\partial_\tau{A}|/c$, then we can neglect the $(\partial_\xi A)^2$ term.
Making the order-of-magnitude estimate $\partial_\tau \sim \omega_L/2\pi$ and defining $L_\xi$ as the typical propagation distance over which the field shape changes substantially, this condition becomes equivalent to $\lambda \ll L_\xi$, where $\lambda =2\pi c/ \omega_L$ is the incident laser wavelength.
In other words, if the field evolves over spatial scales which are large compared to the laser wavelength, then the second term of Lagrangian \eqref{eq:Mod_mov_Lem} is negligible.
In this case, the Lagrangians become
\begin{subequations}\label{eq:Model_moving_back}
\begin{align} \label{eq:Mod_mov_Lp_back}
& \mathcal{L}_\mathrm{P} = 4\pi \int \rho \left[ \frac{|\partial_\tau{\bx}|^2}{2} - V(\bx) - \partial_\tau{x}_{\rm e} A \right] \mathrm{d}\mu {\rm d}\xi, \\ \label{eq:Mod_mov_Lem_back}
& \mathcal{L}_\mathrm{EM} = c \int \partial_\tau{A}\,\partial_\xi A  {\rm d}\xi,
\end{align}
\end{subequations}
where the particle Lagrangian is unchanged.
The equations of motion become
\begin{subequations}\label{eq:EOM3} 
\begin{align}\label{eq:EOM3_particle}
& \partial^2_\tau \bx = - \nabla_\bx V(\bx) + \partial_\tau A \hat{\bf x}, \\ \label{eq:EOM3_wave}
& -\partial_\xi \partial_\tau {A} = \frac{2 \pi \rho}{c} \partial_\tau{\overline{x}}_{\rm e}. 
\end{align}
\end{subequations}
Now, the field evolution equation \eqref{eq:EOM3_wave} is only first order in $\xi$.
While it is still technically a second order equation for $A$, this equation can be seen as a first order equation for the electric field in the moving frame ${\cal E}=-\partial_\tau{A}$.
In fact, ${\cal E}$ is the only quantity that appears in the electron equations of motion \eqref{eq:EOM3_particle}, so we obtain a well-posed set of equations for $\bx$ and ${\cal E}$.
The model given by Eqs.~\eqref{eq:Model_moving_back} and \eqref{eq:EOM3} incorporates assumptions HP \ref{h:SAE}-\ref{h:planeP} and HF \ref{h:plane}-\ref{h:back} and describes the propagation of a solely forward-propagating electromagnetic wave through a classical atomic gas.
In Sec.~\ref{sec:HamClass}, we will show that the assumption $ \lambda \ll L_\xi$ is equivalent to assuming that backward-propagating waves are negligible (HF \ref{h:back}), making this a unidirectional approximation \cite{Kole04}.

\subsubsection{One-dimensional electron dynamics (HP \ref{h:SAE}-\ref{h:ePS} and HF \ref{h:plane}-\ref{h:back})}
A model with a reduced electron phase space may be obtained trivially.
For example, the final hypothesis HP \ref{h:ePS} may be implemented by assuming $\bx = x_{\rm e}(\xi,\tau)\hat{\bf x}$.
Doing so only leads to a modification of the particle Lagrangian $\cal L_{\rm P}$.
The equations of motion stemming from these Lagrangians are equivalent to the model used in Ref.~\cite{Berm18}.
The only remaining step to obtain the model equations is to pass from a Lagrangian description of the particles (in the sense of a Lagrangian description of a fluid) of Eq.~\eqref{eq:EOM2_particle} to an Eulerian description in terms of $f(x_{\rm e},v_x,\tau,\xi)$ \cite{Morr05}. 
$f$ is the phase space probability distribution to find an electron with position $x_{\rm e}$ relative to the ion and velocity $v_x$ at time $\tau$ and position $\xi$ along the gas propagation direction, and it is such that $f(x_{\rm e},v_x,0,\xi)=f_0(x_{\rm e},v_x)$.
Hence, the model equations of Ref.~\cite{Berm18} are obtained:
\begin{subequations}\label{eq:EOM_model} 
\begin{align}\label{eq:EOM_modelL}
& \partial_\tau{f} = -v_x\partial_{x_{\rm e}} f + \left[\partial_{x_{\rm e}} V + {\cal E}(\xi,\tau) \right] \partial_{v_x} f, \\ \label{eq:EOM_modelE}
& \partial_\xi {\cal E}  = \frac{2 \pi \rho}{c} \overline{v}_x(\xi,\tau).
\end{align}
\end{subequations}

\subsubsection{Focusing effects}
Lastly, we consider the possibility of reintroducing some of the focusing effects which are manifestly absent from the 1D wave equation.
As mentioned earlier, the on-axis electric field of the focused laser pulse has a $z$-dependent (equivalently, $\xi$-dependent) maximum amplitude and phase, even in the vacuum.
For example, it may be of the form
\begin{equation}\label{eq:E0}
{\cal E}_0(\xi,\tau) = a(\xi)g(\tau)\cos\left(\omega\tau + \phi(\xi)\right).
\end{equation}
Here, $a$ is the $\xi$-dependent amplitude, $\phi$ is the $\xi$-dependent phase, and $g$ is the temporal envelope.
Notably, this type of solution is precluded by the 1D model in vacuum \cite{Kim02}, because in that case, Eq.~\eqref{eq:EOM3_wave} gives $-\partial_\xi \partial_\tau{A}_0 = \partial_\xi {\cal E}_0 = 0$, where $A_0$ is the on-axis vector potential of the focused laser pulse in vacuum.
Therefore, we are not able to incorporate such effects self-consistently.
However, we can incorporate them in an external fashion.

The idea is to let $A$ represent only the radiation generated by the particles, while the incident laser pulse is treated as a given external field $A_0$.
$A_0(\xi,\tau)$ (or $A_0(z,t)$ in the static frame) should be calculated by first solving Maxwell's equations in vacuum for a focused laser pulse, and then evaluating the resulting vector potential on-axis, i.e.\ at $\br_\perp = 0$.
By following this procedure, one would obtain an on-axis electric field like Eq.~\eqref{eq:E0} from the relation ${\cal E}_0 = -\partial_\tau A_0$.
Because Maxwell's equations are linear, the total vector potential is then given by $A_0 + A$.
As such, the only necessary modification to the Lagrangian is to add an $A_0$ term to $\cal L_{\rm P}$ so that it reads
\begin{equation*}
\mathcal{L}_\mathrm{P} = 4\pi \int \rho \left[ \frac{|\dot{\bx}|^2}{2} - V(\bx) - \dot{x}_e (A + A_0) \right] {\rm d}\mu  {\rm d}z.
\end{equation*}
Consequently, Eq.~\eqref{eq:EOM2_particle} would be modified by the addition of $\dot{A}_0\hat{\bf x}$ on the right-hand side, while Eq.~\eqref{eq:EOM3_wave} would be unchanged.
Additionally, one would need to subject $\dot{A}$ to an initial condition that reflects that the radiation produced by the particles is zero before they are reached by the incident laser pulse.

\subsection{Hamiltonian formulation}\label{sec:HamClass}
The derivation of the sequence of reduced models may also be performed using a Hamiltonian formulation.
While this approach is generally more involved, it possesses some advantages over the Lagrangian derivation.
Namely, the Hamiltonian formulation directly yields the equations of motion of the system as a dynamical system, i.e.\ a coupled set of differential equations which are first order in the evolution parameter $t$.
In contrast, the Lagrangian formulation may produce equations which are second order in $t$, and it may even produce equations with multiple evolution parameters.
For example, all the equations of motion in Sec.~\ref{sec:LagClassical} before Eq.~\eqref{eq:EOM_Amoving} are second order in $t$.
Meanwhile, in Eqs.~\eqref{eq:EOM_model}, there are two evolution parameters: $\tau$ is the evolution parameter for Eq.~\eqref{eq:EOM_modelL}, while $\xi$ is the evolution parameter for Eq.~\eqref{eq:EOM_modelE}.
The Hamiltonian formulation also provides a natural way of identifying conserved quantities of the reduced models, including Casimir invariants.

In the following, we will specify the Hamiltonian structure of the classical parent model [Eqs.~\eqref{eq:Maxwell} and \eqref{eq:ClassicalParent}] and implement the hypotheses in Table \ref{tab:Hyp} to obtain the sequence of reduced Hamiltonian models corresponding to those derived in the previous section.
The model equations obtained from the Hamiltonian framework will thus be completely equivalent to those obtained from the Lagrangian framework.

We use the electron probability distribution function $f(\bx,\bp,\br)$ as the particle dynamical variable, where $\bp=(p_x,p_y,p_z)$ is the canonical momentum of the electron.
It is normalized such that $\int f(\bx,\bp,\br) {\rm d} \mu = 1$, where here ${\rm d} \mu = {\rm d}^3 \bx {\rm d}^3 \bp$.
The field dynamical variables are $\bE(\br)$ and $\bA(\br)$.
Observables are thus functionals ${\cal F} = {\cal F}[f(\bx,\bp,\br),\bE(\br),\bA(\br)]$.
We have omitted the implicit time-dependence of the field variables $f$, $\bE$, and $\bA$.
In analogy with the Vlasov-Maxwell system \cite{Mars82}, the parent model Hamiltonian and non-canonical Poisson bracket are
\begin{subequations}\label{eq:ParentHamSys}
\begin{align} \label{eq:ParentHamiltonian}
& {\cal H}[f,\bE,\bA] = {\cal H}_{\rm P} + {\cal H}_{\rm EM}, \\ \label{eq:ParentHamiltonianP}
& {\cal H}_{\rm P}[f,\bA] = \int \rho \,f \left[\frac{1}{2}\left|\bp + \bA\right|^2 + V(\bx)\right] \mathrm{d}\mu {\rm d}^3 \br, \\ \label{eq:ParentHamiltonianEM}
& {\cal H}_{\rm EM}[\bE,\bA] =  \frac{1}{8 \pi} \int \left( |\bE|^2 + c^2 |\nabla \times \bA |^2\right) {\rm d}^3 \br, \\ \label{eq:ParentBracket}
& \{\FF,\GG\} =  \int \left\{ \rho^{-1} \int f \left[\FF_f,\GG_f \right] \mathrm{d} \mu + 4\pi \left( \FF_\bE \cdot \GG_\bA  - \FF_\bA \cdot \GG_\bE \right) \right\} \mathrm{d}^3 \br.
\end{align}
\end{subequations}
We have introduced the canonical Poisson bracket notation $[f,g] = \partial_\bx f \cdot \partial_\bp g - \partial_\bp f \cdot \partial_\bx g$. 
Like the Lagrangian, Hamiltonian \eqref{eq:ParentHamiltonian} is split into a particle Hamiltonian $\HH_{\cal P}$ and an electromagnetic Hamiltonian $\HH_{\rm EM}$, a splitting which will likewise carry over to each of the reduced models in the hierarchy.
Physically, $\HH_{\rm P}$ is the energy of the atomic electrons --kinetic plus potential-- while $\HH_{\rm EM}$ is the energy of the electromagnetic field.	

The equations of motion are obtained using the observable evolution law $\dot{\FF} = \{\FF,\HH\}$.
They are as follows:
\begin{subequations}\label{eq:EOM_ParentHamiltonian}
\begin{align}\label{eq:EOM_Liouville}
\dot{f} & = -(\bp + \bA)\cdot \nabla_\bx f + \nabla_\bx V \cdot \partial_\bp f, \\ \label{eq:EOM_Ampere_Ham}
\dot{\bE} & = c^2 \nabla \times (\nabla \times \bA) + 4\pi 
\rho (\overline{\bp} + \bA), \\ \label{eq:EOM_A}
\dot{\bA} & = -\bE.
\end{align}
\end{subequations}
In Eq.~\eqref{eq:EOM_Ampere_Ham}, we have used the normalization of the distribution function and introduced the ensemble average $\overline{\bp}(\br) = \int \bp f(\bx,\bp,\br)  {\rm d} \mu$.
This system is completely equivalent to Eqs.~\eqref{eq:Maxwell} and \eqref{eq:ClassicalParent}.
Here, $f$ provides an Eulerian description of the particles corresponding to the Lagrangian description used in Eqs.~\eqref{eq:ClassicalParent}.
The substitutions $\overline{\bv} = \overline{\bp} + \bA$ and $\bB = \nabla \times \bA$ make the equivalence of Eq.~\eqref{eq:EOM_Ampere_Ham} and Eq.~\eqref{eq:MaxAmpere} apparent, while taking the curl of Eq.~\eqref{eq:EOM_A} makes the equivalence to \eqref{eq:MaxLenz} apparent.
As before, Eq.~\eqref{eq:MaxDivB} is guaranteed by the definition of $\bA$.

To obtain Gauss' Law, Eq.~\eqref{eq:MaxGauss}, one needs to consider the conserved quantities of this system.
Because the parent model equations \eqref{eq:EOM_ParentHamiltonian} have a Hamiltonian structure given by \eqref{eq:ParentHamSys}, conserved quantities may be found by searching for observables $\FF$ which Poisson commute with the Hamiltonian, i.e.\ $\{\FF,\HH\}=0$.
Thus, $\HH$, the total energy of the system, is conserved.
Gauss' Law is found by realizing that ${\cal C}(\br') = \nabla \cdot [\bE(\br') - 4\pi \rho \overline{\bx}(\br')]$ is also a conserved quantity.
Therefore,  Eq.~\eqref{eq:MaxGauss} is satisfied for all times if it is satisfied initially.
There is also a family of global Casimir invariants of Poisson bracket \eqref{eq:ParentBracket} - that is, a family of observables ${\cal R}$ for which $\{ {\cal R},\FF \}=0$ for any observable $\FF$.
Consequently, Casimir invariants are conserved regardless of the Hamiltonian.
This family is of the form ${\cal R}[f] = \int R(f) {\rm d}\mu {\rm d}^3\br$ \cite{Ye92}, for arbitrary scalar functions $R$, and it is associated with the relabeling symmetry \cite{Padh96}.

\subsubsection{From canonical momentum to velocity}
As a first step in the derivation, we make the standard transformation from canonical momentum $\bp$ to velocity $\bv = (v_x,v_y,v_z)$, by introducing the change of coordinates on the distribution function (see Ref.~\cite{Mars82} for more details)
\begin{equation}\label{eq:fMom}
\widetilde{f}(\bx,\bv,\br) = f(\bx,\bv - \bA(\br),\br).
\end{equation}
The new observables are defined in terms of the old observables as $\widetilde{\FF}[\widetilde{f},\bE,\bA] = \FF [f,\bE,\bA]$.
By using the chain rule, we obtain relations between the functional derivatives of the old observables and those of the new observables.
They are
\begin{equation*}
\FF_f = \widetilde{\FF}_{\widetilde{f}}, \quad \FF_\bE = \widetilde{\FF}_\bE,\quad \FF_\bA = \widetilde{\FF}_\bA + \int \widetilde{f} \partial_\bv \widetilde{\FF}_{\widetilde{f}} {\rm d}\mu.
\end{equation*}
This leads to the particle Hamiltonian and bracket
\begin{subequations}
\begin{align} \label{eq:HamVelocity}
& {\cal H}_{\rm P}[f] = \int \rho \,f \left[\frac{|\bv|^2}{2} + V(\bx)\right] \mathrm{d}\mu {\rm d}^3 \br, \\ 
& \{\FF,\GG\} =  \int \bigg\{ \rho^{-1} \int f \left[\FF_f,\GG_f \right] \mathrm{d} \mu + 4\pi \left( \FF_\bE \cdot \GG_\bA  - \FF_\bA \cdot \GG_\bE \right) \nonumber \\ \label{eq:BrackVelocity}
& + 4\pi \int f \left( \FF_\bE \cdot \partial_\bv \GG_f  - \partial_\bv \FF_f \cdot \GG_\bE \right) {\rm d}\mu \bigg\} \mathrm{d}^3 \br,
\end{align}
\end{subequations}
where the tildes have been neglected for notational simplicity.
Meanwhile, $\HH_{\rm EM}$ remains as Eq.~\eqref{eq:ParentHamiltonianEM}.
Note that, now, ${\rm d}\mu = {\rm d}^3 \bx {\rm d}^3 \bv$.
We also note that for bracket \eqref{eq:BrackVelocity}, the family of Casimirs ${\cal R}$ becomes slightly restricted to only allow scalar functions $R$ that satisfy $R(0)=0$.
This family of Casimirs persists in this form for all of the reduced models which follow.
So far, no approximations have been made.

\subsubsection{Plane-wave fields (HP \ref{h:SAE}-\ref{h:planeP}, HF \ref{h:plane})}
Next, we implement hypotheses HF~\ref{h:plane} and HP~\ref{h:planeP}, that is, we restrict the fields' macroscopic spatial dependence to be on $z$ only.
We assume
\begin{subequations}
\begin{align*}
& \rho = \rho(z), \\
& f = f(\bx,\bv,z), \\
& \bE = \bE(z), \\
& \bA = \bA_\perp(z).
\end{align*}
\end{subequations}
Here, we have also assumed that $A_z=0$ merely for convenience, since it will no longer appear in the Hamiltonian due to HF \ref{h:plane}.
Thus, we may consider restricting our model to the subset of observables $\FF[f,\bE,\bA_\perp]$ which do not depend on $A_z$.
This subset of observables forms a Poisson subalgebra of the algebra of observables under bracket \eqref{eq:BrackVelocity}.
That is, for two observables $\FF$ and $\GG$ which do not depend on $A_z$, $\{\FF,\GG\}$ is also an observable which does not depend on $A_z$.
As such, we are able to restrict our analysis to this subalgebra, meaning we are free to neglect the functional derivatives with respect to $A_z$ in bracket \eqref{eq:BrackVelocity}.
The electromagnetic Hamiltonian and bracket become
\begin{subequations}\label{eq:HamSysPlane}
\begin{align} \label{eq:HamPlane}
& {\cal H}_{\rm EM}[\bE,\bA_\perp] = \frac{1}{8 \pi} \int \left( |\bE|^2 + c^2 |\partial_z \bA_\perp |^2\right) {\rm d} z, \\ 
& \{\FF,\GG\} =  \int \bigg\{ \rho^{-1} \int f \left[\FF_f,\GG_f \right] \mathrm{d} \mu + 4\pi \left( \FF_{\bE} \cdot \GG_{\bA_\perp}  - \FF_{\bA_\perp} \cdot \GG_{\bE} \right) \nonumber \\ \label{eq:BrackPlane}
& + 4\pi \int f \left( \FF_\bE \cdot \partial_\bv \GG_f  - \partial_\bv \FF_f \cdot \GG_\bE \right) {\rm d}\mu \bigg\} \mathrm{d} z,
\end{align}
\end{subequations}
while the particle Hamiltonian is that of Eq.~\eqref{eq:HamVelocity} with ${\rm d}^3 \br $ replaced with ${\rm d} z$.
The equations of motion at this stage become
\begin{subequations}\label{eq:EOM_H1}
\begin{align}\label{eq:EOM_Liouville1}
& \dot{f} = -\bv \cdot \nabla_\bx f + \left( \nabla_\bx V + \bE \right) \cdot \partial_\bv f, \\ \label{eq:EOM_Ampere_Ham1}
& \dot{\bE} = - c^2 \partial^2_{z} \bA_\perp + 4\pi 
\rho \overline{\bv}, \\ \label{eq:EOM_A1}
& \dot{\bA}_\perp = -\bE_\perp, 
\end{align}
\end{subequations}
while the conserved quantity associated with ${\cal C}$ becomes ${\cal C}(z') = \partial_z [E_z(z') - 4\pi \rho \overline{z}_{\rm e}(z')]$.
For this system, it turns out that ${\cal C}$ is conserved because its primitive, $\widetilde{\cal C}(z') = E_z(z')-4\pi\rho \overline{z}_{\rm e}(z')$ is a Casimir invariant of bracket \eqref{eq:BrackPlane}.
Thus, the longitudinal electric field is simply obtained from the longitudinal component of the microscopic dipole moment.
An additional global pair of conserved quantities is created by HP \ref{h:planeP} and HF \ref{h:plane}, given by ${\bf {\cal Q}}_\perp = \int (\bE_\perp - 4\pi\rho \overline{\bx}_\perp ){\rm d} z$.

\subsubsection{Transverse, linearly-polarized fields (HP \ref{h:SAE}-\ref{h:planeP} and HF \ref{h:plane}-\ref{h:LP})}
To implement HP~\ref{h:electrostatic}, i.e. the assumption that $E_z$ is negligible, we drop the $E_z^2$ term from the Hamiltonian.
Thus, ${\cal H}$ no longer depends on $E_z$, and we may consider restricting our model to the subset of observables $\FF[f,\bE_\perp,\bA_\perp]$ which do not depend on $E_z$.
This subset of observables forms a Poisson subalgebra of the algebra of observables under bracket \eqref{eq:BrackPlane}.
That is, for two observables $\FF$ and $\GG$ which do not depend on $E_z$, $\{\FF,\GG\}$ is also an observable which does not depend on $E_z$.
As such, we are able to restrict our analysis to this subalgebra, meaning we are free to neglect the functional derivatives with respect to $E_z$ in bracket \eqref{eq:BrackPlane}.
In effect, the new bracket is \eqref{eq:BrackPlane} with $\bE$ replaced by $\bE_\perp$.
Henceforth, we will also invoke hypothesis HF \ref{h:LP}, so we shall drop the $E_y$ and $A_y$ terms from the Hamiltonian.
They may also be removed from the bracket with a subalgebra argument.
The electromagnetic Hamiltonian and bracket for the system under hypotheses HP \ref{h:SAE}-\ref{h:planeP} and HF \ref{h:plane}-\ref{h:LP} become
\begin{subequations}\label{eq:HamSysLP}
\begin{align} \label{eq:HamLP}
& {\cal H}_{\rm EM}[E,A] = \frac{1}{8 \pi}  \int \left( E^2 + c^2 (\partial_z A)^2\right) {\rm d} z, \\ 
& \{\FF,\GG\} =  \int \bigg\{ \rho^{-1} \int f \left[\FF_f,\GG_f \right] \mathrm{d} \mu + 4\pi \left( \FF_{E} \GG_A  - \FF_A \GG_E \right) \nonumber \\ \label{eq:BrackLP}
& + 4\pi \int f \left( \FF_E  \partial_{v_x} \GG_f  - \partial_{v_x} \FF_f \GG_E \right) {\rm d}\mu \bigg\} \mathrm{d} z,
\end{align}
\end{subequations}
where we have dropped the $x$ subscripts on the electromagnetic fields and $\HH_{\rm P}$ remains unchanged.
Now, the equations of motion are
\begin{subequations}\label{eq:EOM_H2}
\begin{align}\label{eq:EOM_Liouville2}
& \dot{f} = -\bv \cdot \nabla_\bx f + \left( \nabla_\bx V + E \hat{\bx} \right) \cdot \partial_\bv f, \\ \label{eq:EOM_Ampere_Ham2}
& \dot{E} = - c^2 \partial^2_{z} A + 4\pi \rho \overline{v}_x, \\ \label{eq:EOM_A2}
& \dot{A} = -E.
\end{align}
\end{subequations}

\subsubsection{Forward- and backward- propagating waves}
Before implementing hypothesis HF \ref{h:back}, we perform a reduction on the field variables which elucidates the natural separation of the electromagnetic field into a forward- and backward-propagating wave.
The reduction is given by
\begin{subequations}\label{eq:waves}
\begin{align}
& \alpha = \frac{1}{2} \left(E + c \partial_z A \right), \\
& \beta = \frac{1}{2} \left(E - c \partial_z A \right),
\end{align}
\end{subequations}
where $\alpha$ is the forward-propagating field and $\beta$ is the backward-propagating field.
The electric field is simply expressed in these reduced variables as $E = \alpha+\beta$.
The functional derivatives appearing in bracket \eqref{eq:BrackLP} are obtained in terms of the new variables using the chain rule:
\begin{subequations}
\begin{align*}
\FF_E & = \frac{1}{2}\left(\widetilde{\FF}_\alpha + \widetilde{\FF}_\beta \right), \\
\FF_A & = -\frac{c}{2}\left(\partial_z \widetilde{\FF}_\alpha - \partial_z \widetilde{\FF}_\beta \right).
\end{align*}
\end{subequations}
The electromagnetic Hamiltonian and bracket become
\begin{subequations}\label{eq:HamSysWave}
\begin{align} \label{eq:HamWave}
& {\cal H}_{\rm EM}[\alpha,\beta] = \frac{1}{4 \pi} \int  \left[ \alpha^2 + \beta^2 \right] {\rm d} z, \\ 
& \{\FF,\GG\} =  \int \bigg\{ \rho^{-1} \int f \left[\FF_f,\GG_f \right] \mathrm{d} \mu - 2\pi c \left( \FF_{\alpha} \partial_z \GG_\alpha  - \FF_\beta \partial_z \GG_\beta \right) \nonumber \\ \label{eq:BrackWave}
& + 2\pi \int f \left( \FF_\alpha  \partial_{v_x} \GG_f  - \partial_{v_x} \FF_f \GG_\alpha + \FF_\beta  \partial_{v_x} \GG_f  - \partial_{v_x} \FF_f \GG_\beta \right) {\rm d}\mu \bigg\} \mathrm{d} z,
\end{align}
\end{subequations}
while ${\cal H}_{\rm P}$ is unaffected.
In these coordinates, the equations of motion are
\begin{subequations}\label{eq:EOM_H3}
\begin{align}\label{eq:EOM_Liouville3}
\dot{f} & = -\bv \cdot \nabla_\bx f + \left[ \nabla_\bx V + (\alpha + \beta) \hat{\bx} \right] \cdot \partial_\bv f, \\ \label{eq:EOM_alpha}
\dot{\alpha} & = -c \partial_z \alpha + 2\pi \rho \overline{v}_x, \\ \label{eq:EOM_beta}
\dot{\beta} & = c \partial_z \beta + 2\pi \rho \overline{v}_x.
\end{align}
\end{subequations}
From these equations, it is clear that $\alpha$ is the forward-propagating part of the electromagnetic field and $\beta$ is the backward-propagating part.
Indeed, in vacuum ($\rho = 0$), the solution of Eq.~\eqref{eq:EOM_alpha} would be $\alpha(z,t) = \alpha_0(z-ct)$, where $\alpha(z,0) = \alpha_0(z)$, and the solution of equation Eq.~\eqref{eq:EOM_beta} would be $\beta(z,t) = \beta_0(z+ct)$, where $\beta(z,0) = \beta_0(z)$.

Equations \eqref{eq:waves} do not constitute a change of variables because it is not possible to determine $A$ uniquely from $\alpha$ and $\beta$; it is only determined up to a constant.
This constant has no physical significance because Hamiltonian \eqref{eq:HamLP} only depends on $\partial_z A$.
In fact, it is a manifestation of the gauge freedom inherent to the potential description of the magnetic field.
Due to the reduction to $\alpha$ and $\beta$, which eliminates the remaining gauge freedom, two global Casimir invariants of bracket \eqref{eq:BrackWave} are created.
They are related to ${\cal Q}_\perp$, a conserved quantity of the previous system, and are given by
\begin{subequations}\label{eq:casimirs}
\begin{align}
& {\cal Q}_\alpha = \int (\alpha - 2\pi\rho \overline{x}_{\rm e} ){\rm d} z, \\
& {\cal Q}_\beta = \int (\beta - 2\pi\rho \overline{x}_{\rm e} ){\rm d} z.
\end{align}
\end{subequations}
By direct calculation, it is possible to verify that bracket \eqref{eq:BrackWave} satisfies the Jacobi identity, so it is a genuine Poisson bracket and this reduction has preserved the Hamiltonian structure of the previous model.

\subsubsection{Unidirectional approximation (HP \ref{h:SAE}-\ref{h:planeP} and HF \ref{h:plane}-\ref{h:back})}
Having clearly separated the forward- and backward- propagating parts of the electromagnetic field, it becomes straightforward to make the unidirectional approximation and remove the backward-propagating part (HF \ref{h:back}).
If $\beta$ is assumed to be small, then the $\beta^2$ term may be neglected from Hamiltonian \eqref{eq:HamWave}.
Then, the Hamiltonian no longer depends on $\beta$, and it happens that observables $\FF[f,\alpha]$ which do not depend on $\beta$ form a Poisson subalgebra under bracket \eqref{eq:BrackWave}.
This leaves the electromagnetic Hamiltonian and bracket for the system incorporating hypotheses HP \ref{h:SAE}-\ref{h:planeP} and HF \ref{h:plane}-\ref{h:back}:
\begin{subequations}\label{eq:HamSysForward}
\begin{align} \label{eq:HamForward}
& {\cal H}_{\rm EM}[\alpha] = \int \frac{\alpha^2}{4 \pi} {\rm d} z, \\ 
& \{\FF,\GG\} =  \int \bigg\{ \rho^{-1} \int f \left[\FF_f,\GG_f \right] \mathrm{d} \mu - 2\pi c \FF_{\alpha} \partial_z \GG_\alpha \nonumber \\ \label{eq:BrackForward}
& + 2\pi \int f \left( \FF_\alpha  \partial_{v_x} \GG_f  - \partial_{v_x} \FF_f \GG_\alpha \right) {\rm d}\mu \bigg\} \mathrm{d} z.
\end{align}
\end{subequations}
The equations of motion are Eqs.~\eqref{eq:EOM_Liouville3} and \eqref{eq:EOM_alpha}, with $\beta=0$.

Now, we are able to justify that neglecting backward-propagating waves is equivalent to assuming $\lambda \ll L_\xi$, which is used in Sec.~\ref{sec:LagUni} to implement HF \ref{h:back}.
In the unidirectional approximation, the electric field becomes $E=\alpha$.
Thus, Eq.~\eqref{eq:EOM_alpha} is the evolution equation for the electric field of the laser.
Comparing this equation to the corresponding one from the Lagrangian derivation, Eq.~\eqref{eq:EOM3_wave}, we find that they are completely equivalent.
The equivalence may be seen either by moving Eq.~\eqref{eq:EOM3_wave} back to the rest frame, or by moving Eq.~\eqref{eq:EOM_alpha} to the moving frame [${\cal E}(\xi,\tau) = \alpha(\xi,\tau+\xi/c)$] and recalling that $E = -\dot{A}$.
Each of these equations is obtained under seemingly unrelated hypotheses: that $\beta$ is negligible for Eq.~\eqref{eq:EOM_alpha}, versus that $\lambda \ll L_\xi$ for Eq.~\eqref{eq:EOM3_wave}.
Because the field evolution equations resulting from each hypothesis are equivalent, we conclude that the hypotheses are in fact equivalent.
Therefore, the unidirectional approximation is the same as assuming the laser field evolves over large length scales compared to the incident laser wavelength.

\subsubsection{One-dimensional electron dynamics (HP \ref{h:SAE}-\ref{h:ePS} and HF \ref{h:plane}-\ref{h:back})}
As in the Lagrangian case, a reduced electron phase space model (HP \ref{h:ePS}) is straightforward to implement.
One simply considers a distribution function on a lower-dimensional phase space.
For example, for one-dimensional electron motion, one assumes $f = f(x_{\rm e},v_x,z)$, with the obvious modifications to Eqs.~\eqref{eq:HamSysForward}.

\subsubsection{Focusing effects}
For incorporating focusing effects, the procedure is also similar to the Lagrangian case.
The dynamical field $\alpha$ is considered to be the radiation generated solely by the particles.
Meanwhile, the incident laser radiation is taken to be a given external, time-dependent field $E_0(z,t)$.
Then, one only needs to modify the particle Hamiltonian \eqref{eq:HamVelocity} such that it reads
\begin{equation}\label{eq:HamFocus}
{\cal H}_{\rm P}[f,t] = \int \rho  \,f \left[\frac{|\bv|^2}{2} + V(\bx) + x_{\rm e} E_0(z,t)  \right] \mathrm{d}\mu {\rm d} z,
\end{equation}
while $\HH_{\rm EM}$ remains Eq.~\eqref{eq:HamForward} and the Poisson bracket remains Eq.~\eqref{eq:BrackForward}.
Note that Hamiltonian \eqref{eq:HamFocus} is time-dependent, so we must expand our set of observables to allow observables of the type $\FF[f,\alpha,t]$.
Further, we must redefine the evolution law as $\dot{\FF} = \{\FF,\HH\} + \partial_t \FF$.
Consequently, total Hamiltonian $\HH = \HH_{\rm P} + \HH_{\rm EM}$ is no longer a conserved quantity.
However, ${\cal Q}_\alpha$ remains a conserved quantity (because $\partial_t {\cal Q}_\alpha = 0$), and it may be restored to the status of Casimir invariant by autonomizing the system and extending the Poisson bracket appropriately.
We remark that focusing effects may also be incorporated in this way to each of the previous reduced models in the hierarchy by adding the focusing field to the model's particle Hamiltonian (and using the corresponding bracket).
Therefore, ${\cal Q}_\beta$ and $\widetilde{\cal C}(z')$ are also conserved in the presence of a time-dependent external field.

\section{Quantum reduced models}\label{sec:quantum}
Each of the reduced models derived in the previous section has a quantum analog.
These models may be found by sequentially applying the assumptions of Table \ref{tab:Hyp} to the parent quantum model, Eqs.~\eqref{eq:Maxwell} and \eqref{eq:QuantParent}, and this may be also accomplished using a variational formulation.
Due to the similarity between the variational formulations of the classical and quantum models and our focus on reducing the degrees of freedom associated with the electromagnetic field, the derivation of the quantum models is nearly identical to that of the classical models.
Hence, we give few details on the calculations and focus on the results.

\subsection{Lagrangian formulation}
The action for the parent quantum model is given by \[ {\cal A}[\psi(\bx,t;\br),\psi^*(\bx,t;\br),\varphi(\br,t),\bA(\br,t)] = \int ({\cal L}_{\rm P} + {\cal L}_{\rm EM})\mathrm{d}t.\]
The electron displacement field $\bx(\br,t;\bx_0,\bv_0)$ is replaced by the wave function $\psi$ and its complex conjugate, $\psi^*$.
Here, and in each of the models of the hierarchy, ${\cal L}_{\rm EM}$ will be the same as in the corresponding classical model.
Meanwhile, for the parent quantum model, ${\cal L}_{\rm P}$ is given by
\begin{equation}\label{eq:LagParentQ}
{\cal L}_{\rm P} = 4\pi \int \rho \left\{  i \psi^* \dot{\psi} - \psi^* \left[ - \frac{1}{2}\nabla_\bx^2  + V(\bx) - \bx \cdot (\nabla \varphi + \dot{\bA} ) \right]\psi \right\} \mathrm{d}^3 \bx \mathrm{d}^3 \br.
\end{equation}
Recognizing the appearance of the Hamiltonian operator  $\hat{H} = -\nabla_\bx^2/2 + V(\bx) + \bE \cdot \bx$ in Eq.~\eqref{eq:LagParentQ}, it is clear that ${\cal L}_{\rm P}$ has a phase-space Lagrangian form \cite{Masi05}.
Imposing ${\cal A}_{\psi^*} = 0$ yields Schr\"odinger equation \eqref{eq:ParentSchrod} (with its complex conjugate for ${\cal A}_{\psi}=0)$, while setting the variations with respect to the potentials to zero yields
\begin{subequations}\label{eq:EOM_qparent}
\begin{align}
& -(\nabla^2 \varphi + \nabla\cdot \dot{\bA}) = 4\pi \nabla \cdot \left(\rho \int \bx \psi^* \psi\, \mathrm{d}^3\bx\right), \\ \label{eq:EOM_qparent_amp}
& -(\nabla \dot{\varphi} + \ddot{\bA}) = c^2 \nabla \times (\nabla \times \bA) + 4 \pi \rho\, \partial_t \left(\int \bx \psi^* \psi \,\mathrm{d}^3\bx \right) .
\end{align}
\end{subequations}
The integrals on the right-hand sides of Eqs.~\eqref{eq:EOM_qparent} are clearly recognized as $\overline{\bx}$, as defined in Eq.~\eqref{eq:ParentExpecVal}. 
With this and the definition of the potentials in mind, we confirm the correspondence with Eqs.~\eqref{eq:Maxwell}.

The first model in the hierarchy is obtained by applying hypotheses HP \ref{h:planeP} and HF \ref{h:plane}.
These hypotheses are summed up by Eqs.~\eqref{eq:fieldAssump}, with Eq.~\eqref{eq:fieldAssump_x} replaced by
\begin{equation*}
\psi = \psi(\bx,t;z).
\end{equation*}
The particle Lagrangian becomes
\begin{equation*}
{\cal L}_{\rm P} = 4\pi \int \rho \left\{  i \psi^* \dot{\psi} - \psi^* \left[ - \frac{1}{2}\nabla_\bx^2  + V(\bx) - \bx \cdot (\partial_z \varphi \hat{\bf z} + \dot{\bA}_\perp ) \right]\psi \right\} \mathrm{d}^3 \bx {\rm d} z,
\end{equation*}
and the Schr\"odinger equation becomes
\begin{equation*}
i \dot{\psi} = - \frac{1}{2}\nabla^2_\bx \psi + \big[V(\bx) - \partial_z \varphi(z,t)z_{\rm e} - \dot{\bA}_\perp(z,t) \cdot \bx_\perp \big] \psi.
\end{equation*}
Meanwhile, the field equations remain Eqs.~\eqref{eq:EOM1Gauss} and \eqref{eq:EOM1Wave}, with the ensemble averages of the electron positions computed using the corresponding quantum expectation values.
In general, the field equations at each level of the hierarchy will be the same in the quantum case as in the classical case.

Next, we implement hypothesis HF \ref{h:electrostatic}.
The particle Lagrangian is
\begin{equation*}
{\cal L}_{\rm P} = 4\pi \int \rho \left\{  i \psi^* \dot{\psi} - \psi^* \left[ - \frac{1}{2}\nabla_\bx^2  + V(\bx) - \bx_\perp \cdot \dot{\bA}_\perp \right]\psi \right\} \mathrm{d}^3 \bx {\rm d}z.
\end{equation*}
When also making assumption HF \ref{h:LP} ($A_y=0$), the Schr\"odinger equation becomes
\begin{equation}\label{eq:EOM2_schrod}
i \dot{\psi} = - \frac{1}{2}\nabla^2_\bx \psi + \big[V(\bx) -\dot{A}(z,t) x_{\rm e} \big] \psi.
\end{equation}
The corresponding field equation is Eq.~\eqref{eq:EOM2_wave}.
Eqs.~\eqref{eq:EOM2_schrod} and \eqref{eq:EOM2_wave} are equivalent to the model used in Ref.~\cite{Chri98}.
Now, we go into the moving frame $\xi = z, \tau = t - z/c$.
The wave function in the moving frame $\widetilde{\psi}$ is defined
\begin{equation*}
\widetilde{\psi}(\bx,\tau;\xi) = \psi(\bx,\tau+\xi/c;\xi).
\end{equation*}
As in the classical case, the functional form of ${\cal L}_{\rm P}$ is unchanged by this transformation.
The moving-frame formulation of this model is employed in Ref.~\cite{Shon00}.
Finally, hypothesis HF \ref{h:back} is implemented only on ${\cal L}_{\rm EM}$, just as in the classical model.
The resulting model corresponds to what is referred to as the reduced model of Ref.~\cite{Shon00}.
Adding on hypothesis HP \ref{h:ePS} (taking $\psi = \psi(x_{\rm e},\tau;\xi)$) leads to the particle Lagrangian
\begin{equation}\label{eq:Lag_q_final}
{\cal L}_{\rm P} = 4\pi \int \rho \left\{  i \psi^* \partial_\tau{\psi} - \psi^* \left[ -\frac{1}{2}\partial^2_{x_{\rm e}} + V(x_{\rm e})  - x_{\rm e} \partial_\tau {A} \right] \psi  \right\} \mathrm{d}x_{\rm e} {\rm d} \xi,
\end{equation}
with the corresponding field Lagrangian \eqref{eq:Mod_mov_Lem_back}.
This leads to the quantum model equations that we used in Ref.~\cite{Berm18},
\begin{subequations}\label{eq:EOM_modelq}
\begin{align}
i \partial_\tau{\psi} & = - \frac{1}{2}\partial^2_{x_{\rm e}} \psi + \big[V(x_{\rm e}) +{\cal E}(\xi,\tau) x_{\rm e} \big] \psi, \\ \label{eq:EOM_model_waveq}
\partial_\xi {\cal E}  & = \frac{2 \pi \rho}{c} \overline{v}_x(\xi,\tau),
\end{align}
\end{subequations}
where we have again made the substitution ${\cal E} = -\partial_\tau A$.
Also, in Eq.~\eqref{eq:EOM_model_waveq}, we have used Ehrenfest's theorem, $\overline{v}_x = -i \int \psi^* \partial_{x_{\rm e}} \psi \, {\rm d} x_{\rm e} = \partial_\tau \overline{x}_{\rm e}$.
If desired, focusing effects may be added by adding the appropriate term to Eq.~\eqref{eq:Lag_q_final}, similarly to the classical case.

\subsection{Hamiltonian formulation}
Now we provide the derivation of the quantum reduced model in the Hamiltonian formulation.
In this case, observables are functionals \[{\cal F} = {\cal F}[\psi(\bx;\br),\psi^*(\bx;\br),\bE(\br),\bA(\br)],\]
where we omit the implicit time-dependence of the variables.
The Hamiltonian and bracket of the quantum parent model are
\begin{subequations}
\begin{align} \label{eq:ParentHamiltonianQ}
& {\cal H}[\psi,\psi^*,\bE,\bA] = {\cal H}_{\rm P} + {\cal H}_{\rm EM} \\ \label{eq:ParentHamiltonianQP}
& {\cal H}_{\rm P}[\psi,\psi^*,\bA] = \int \rho  \psi^* \left[\frac{1}{2} \left(-i \nabla_\bx + \bA\right)^2+ V(\bx) \right]\psi\, \mathrm{d}^3 \bx {\rm d}^3 \br, \\ \label{eq:ParentBracketQ}
& \{\FF,\GG\} =  \int \left\{ -i \rho^{-1} \int \left(\FF_\psi \GG_{\psi^*} - \FF_{\psi^*} \GG_{\psi}\right) \mathrm{d}^3 \bx + 4\pi \left( \FF_\bE \cdot \GG_\bA  - \FF_\bA \cdot \GG_\bE \right) \right\} \mathrm{d}^3 \br.
\end{align}
\end{subequations}
The full Hamiltonian ${\cal H}$  has the same splitting into particle and electromagnetic parts, with the electromagnetic part given by Eq.~\eqref{eq:ParentHamiltonianEM}, as in the classical case.
In fact, at every level of the hierarchy, the electromagnetic Hamiltonian of the quantum model will by identical to that of the classical model.
The particle Hamiltonian Eq.~\eqref{eq:ParentHamiltonianQP} is recognized to be the expectation value of the Hamiltonian operator $\hat{H} = (-i\nabla_\bx + \bA)^2/2 + V(\bx)$, integrated over the macroscopic gas.
Thus, the physical meaning of ${\cal H}_{\rm P}$ --the sum of the energies of all the atoms-- is also the same in both the quantum and classical cases.

The equations of motion are obtained from $\dot{\FF} = \{\FF,\HH\}$ which, for the wave function, gives
\begin{equation}\label{eq:EOM_schrod_ham}
\dot{\psi} =  -i\left[\frac{1}{2}(-i\nabla_\bx + \bA)^2\psi + V(\bx)\psi\right],
\end{equation}
while those of the fields are given by Eqs.~\eqref{eq:EOM_Ampere_Ham} and \eqref{eq:EOM_A}, as in the classical case.
Here, the ensemble-averaged canonical momentum is given by $\overline \bp = -i \int \psi^* \nabla_\bx \psi {\rm d}^3 \bx$.
This system also possesses the same conserved quantity $\cal C(\br')$ as in the classical case, with the ensemble average $\overline{\bx}$  computed in the appropriate way.
In fact, each of the conserved quantities of the classical hierarchy of reduced models have an analog in the quantum hierarchy of reduced models.
For instance, here also the quantum Hamiltonian Eq.~\eqref{eq:ParentHamiltonianQ} is conserved.
Additionally, instead of the Casimirs ${\cal R}$, the quantum system conserves the norms of the wave functions, ${\cal N}(\br') = \int \psi^*(\bx;\br') \psi(\bx;\br') {\rm d}^3 \bx$ (though they are not Casimirs of bracket \eqref{eq:ParentBracketQ}).

Equation \eqref{eq:EOM_schrod_ham} is equivalent to Eq.~\eqref{eq:ParentSchrod}, which can be seen by making an appropriate unitary transformation on $\psi$.
This transformation is referred to as going from the velocity gauge to the length gauge in the quantum description \cite{Band13}, and it is the analog of the change from canonical momentum to velocity in the classical derivation (Eq.~\eqref{eq:fMom}).
The change of variables is given by
\begin{equation}\label{eq:unitary}
\widetilde{\psi} = \exp[i \bx \cdot \bA] \psi,
\end{equation}
with the corresponding equation for $\psi^*$.
Using the chain rule, the functional derivatives transform as
\begin{subequations}\label{eq:unitary_f}
\begin{align}
& \FF_\psi = \exp[i \bx \cdot \bA] \widetilde{F}_{\widetilde{\psi}}, \\
& \FF_{\psi^*} = \exp[-i \bx \cdot \bA] \widetilde{F}_{\widetilde{\psi}^*}, \\
& \FF_{\bA} = i\bx \widetilde{\psi}\widetilde{\FF}_{\widetilde{\psi}} - i \bx \widetilde{\psi}^* \FF_{\widetilde{\psi}^*}, \\
& \FF_\bE = \widetilde{\FF}_\bE.
\end{align}
\end{subequations}
Substituting Eqs.~\eqref{eq:unitary} and \eqref{eq:unitary_f} into Hamiltonian \eqref{eq:ParentHamiltonianQP} and bracket \eqref{eq:ParentBracketQ}, respectively, yields
\begin{subequations} \label{eq:HamSysVelQ}
\begin{align} \label{eq:HamVelQ}
& {\cal H}_{\rm P}[\psi,\psi^*] = \int \rho \psi^* \left[-\frac{1}{2} \nabla_\bx^2  + V(\bx)\right] \psi\, \mathrm{d}^3 \bx {\rm d}^3 \br, \\ \label{eq:BracketVelQ}
& \{\FF,\GG\} =  \int \bigg\{ -i \rho^{-1} \int \left(\FF_\psi \GG_{\psi^*} - \FF_{\psi^*} \GG_{\psi}\right) \mathrm{d}^3 \bx + 4\pi \left( \FF_\bE \cdot \GG_\bA  - \FF_\bA \cdot \GG_\bE \right) \nonumber \\
& + 4\pi i \int \left[ \psi \left( \FF_\bE \GG_\psi - \FF_\psi \GG_\bE \right) - \psi^* \left( \FF_\bE \GG_{\psi^*} - \FF_{\psi^*} \GG_\bE \right) \right] \cdot \bx \, {\rm d}^3 \bx  \bigg\} \mathrm{d}^3 \br,
\end{align}
\end{subequations}
where the tildes have been removed.
Now, computing the equation of motion for $\psi$, one obtains Eq.~\eqref{eq:ParentSchrod}.

Deriving the hierarchy of quantum models is straightforward because, as mentioned previously, $\HH_{\rm EM}$ is always the same as in the classical case.
Furthermore, since only $\HH_{\rm EM}$ and the Poisson bracket are modified in deriving the classical reduced models (see Sec.~\ref{sec:HamClass}), the only new information here is the relevant Poisson bracket for each quantum model.
Meanwhile, ${\cal H}_{\rm P}$ is essentially always given by Eq.~\eqref{eq:HamVelQ}.
The first quantum model in the hierarchy, taking into account hypotheses HP \ref{h:SAE}-\ref{h:planeP} and HF \ref{h:plane}, has the following Poisson bracket:
\begin{align}
& \{\FF,\GG\} =  \int \bigg\{ -i \rho^{-1} \int \left(\FF_\psi \GG_{\psi^*} - \FF_{\psi^*} \GG_{\psi}\right) \mathrm{d}^3 \bx + 4\pi \left( \FF_\bE \cdot \GG_{\bA_\perp}  - \FF_{\bA_\perp} \cdot \GG_\bE \right) \nonumber \\ \label{eq:BrackPlaneQ}
& + 4\pi i \int \left[ \psi \left( \FF_\bE \GG_\psi - \FF_\psi \GG_\bE \right) - \psi^* \left( \FF_\bE \GG_{\psi^*} - \FF_{\psi^*} \GG_\bE \right) \right] \cdot \bx \, {\rm d}^3 \bx  \bigg\} \mathrm{d}z.
\end{align}
Like in the classical case, this bracket has $\widetilde{\cal C}$ as a Casimir invariant and ${\bf {\cal Q}}_\perp$ as a conserved quantity.
The corresponding Schr\"odinger equation is unchanged from the parent model, while the field equations are given by Eqs.~\eqref{eq:EOM_Ampere_Ham1} and \eqref{eq:EOM_A1}.
Note that, here, the definition of $\overline{\bv}$ is the same as the definition of $\overline{\bp}$ in the quantum case given previously.
The second quantum model, incorporating also HF \ref{h:electrostatic}, has the same bracket with $\bE$ replaced by $\bE_\perp$.

Adding on HF \ref{h:LP}, the bracket is simplified to
\begin{align}
& \{\FF,\GG\} =  \int \bigg\{ -i \rho^{-1} \int \left(\FF_\psi \GG_{\psi^*} - \FF_{\psi^*} \GG_{\psi}\right) \mathrm{d}^3 \bx + 4\pi \left( \FF_E \GG_{A}  - \FF_{A}\GG_E \right) \nonumber \\ \label{eq:BrackLPQ}
& + 4\pi i \int \left[ \psi \left( \FF_E \GG_\psi - \FF_\psi \GG_E \right) - \psi^* \left( \FF_E \GG_{\psi^*} - \FF_{\psi^*} \GG_E \right) \right]x_{\rm e} \, {\rm d}^3 \bx  \bigg\} \mathrm{d}z.
\end{align}
Performing reduction \eqref{eq:waves} on the field variables, the bracket becomes
\begin{align}
& \{\FF,\GG\} =  \int \bigg\{ -i \rho^{-1} \int \left(\FF_\psi \GG_{\psi^*} - \FF_{\psi^*} \GG_{\psi}\right) \mathrm{d}^3 \bx - 2\pi c \left( \FF_{\alpha} \partial_z \GG_\alpha  - \FF_\beta \partial_z \GG_\beta \right) \nonumber \\ 
& + 2\pi i \int [ \psi \left( \FF_\alpha \GG_\psi - \FF_\psi \GG_\alpha \right) - \psi^* \left( \FF_\alpha \GG_{\psi^*} - \FF_{\psi^*} \GG_\alpha \right)  \nonumber \\ \label{eq:BrackWaveQ}
& + \psi \left( \FF_\beta \GG_\psi - \FF_\psi \GG_\beta \right) - \psi^* \left( \FF_\beta \GG_{\psi^*} - \FF_{\psi^*} \GG_\beta \right)] x_{\rm e} \, {\rm d}^3 \bx  \bigg\} \mathrm{d}z.
\end{align}
Again, the Casimirs \eqref{eq:casimirs} are created by the reduction.
As in the classical case, implementing HF \ref{h:back} consists of removing the terms with functional derivatives with respect to $\beta$ from bracket \eqref{eq:BrackWaveQ}.
Finally, hypothesis HP \ref{h:ePS} is implemented by taking $\psi = \psi(x_{\rm e};z)$.
For reference, the Hamiltonian and bracket of the model taking into account HP \ref{h:SAE}-\ref{h:ePS} and HF \ref{h:plane}-\ref{h:back} are
\begin{subequations}
\begin{align}\label{eq:HamModelQ}
& \HH_{\rm P}[\psi,\psi^*] = \int \rho \psi^* \left[ -\frac{1}{2}  \partial_{x_{\rm e}}^2 + V(x_{\rm e})\right]\psi\, \mathrm{d}x_{\rm e} {\rm d}z, \\
& \{\FF,\GG\} =  \int \bigg\{ -i \rho^{-1} \int \left(\FF_\psi \GG_{\psi^*} - \FF_{\psi^*} \GG_{\psi}\right) \mathrm{d}x_{\rm e} - 2\pi c \FF_{\alpha} \partial_z \GG_\alpha  \nonumber \\
& + 2\pi i \int [ \psi \left( \FF_\alpha \GG_\psi - \FF_\psi \GG_\alpha \right) - \psi^* \left( \FF_\alpha \GG_{\psi^*} - \FF_{\psi^*} \GG_\alpha \right) ] x_{\rm e} \, {\rm d} x_{\rm e}  \bigg\} \mathrm{d}z,
\end{align}
\end{subequations}
with $\HH_{\rm EM}$ given by Eq.~\eqref{eq:HamForward}.
The equations of motion stemming from this Hamiltonian system are equivalent to Eqs.~\eqref{eq:EOM_modelq}.
Focusing effects may be incorporated by adding $E_0(z,t)x_{\rm e}$ to the Hamiltonian operator in Eq.~\eqref{eq:HamModelQ}.

\section{Conclusion}\label{sec:concl}
To summarize, we have presented derivations of a hierarchy of reduced classical and quantum models for the propagation of intense laser pulses in atomic gases.
In particular, we derived the models used in Ref.~\cite{Berm18} to study HHG, and along the way, we derived the reduced quantum models used in Refs.~\cite{Chri98,Shon00}.
By consistently applying simplifying hypotheses within a variational formulation, whether Lagrangian or Hamiltonian, we have ensured that our reduced models preserve the mathematical structure of the parent models.
Using the Hamiltonian formulation, we were able to easily identify conserved quantities of both the classical and quantum systems.
In particular, the conserved quantities $\widetilde{\cal C}(z')$, ${\cal Q}_\alpha$, and ${\cal Q}_\beta$ are interesting because as Casimir invariants, they are conserved even in the presence of a time-dependent external field, unlike the Hamiltonian.
Knowledge of these conservation laws can provide a useful benchmark for numerical codes for solving these model equations.
While we focus on first-principles microscopic models of the atomic response in gases, we anticipate that our methodology can be extended to employ reduced models of the atomic response (e.g. in terms of a macroscopic polarization with an explicit nonlinear dependence on the electric field) which are commonly used in nonlinear optics \cite{Kole04,Kole13}.
Further, variational formulations employing microscopic models of condensed phase systems should also be possible.

\section*{Acknowledgements}
The project leading to this research has received funding from the European Union's Horizon 2020 research and innovation program under the Marie Sk{\l}odowska-Curie grant agreement No 734557. S.A.B. and T.U. acknowledge funding from the NSF (Grant No. PHY1602823). This material is based upon research supported by the Chateaubriand Fellowship of the Office for Science \& Technology of the Embassy of France in the United States. S.A.B. acknowledges funding from the Georgia Tech College of Sciences for extended visits to Marseille.

\end{document}